\documentclass[prb,twocolumn,floatfix,superscriptaddress]{revtex4-2}

\usepackage{amssymb,amsmath,graphicx,afterpage}
\usepackage{color,bm}
\usepackage[colorlinks,bookmarks=false,citecolor=blue,linkcolor=blue,urlcolor=blue]{hyperref}
\usepackage{cancel}
\pdfoutput=1

\begin{document}

\clubpenalty=10000
\widowpenalty=10000
\brokenpenalty=10000
\tolerance=9000
\hyphenpenalty=10000

%Giving a title

\title{\boldmath Magnetoelectric spectroscopy of spin excitations in LiCoPO$_4$\unboldmath}

%Giving the authors and affiliation
\author{V. Kocsis}
\affiliation{RIKEN Center for Emergent Matter Science (CEMS), Wako, Saitama 351-0198, Japan}
\affiliation{Department of Physics, Budapest University of Technology and Economics and MTA-BME Lend\"ulet Magneto-optical Spectroscopy Research Group, 1111 Budapest, Hungary}

\author{S. Bord\'acs}
\affiliation{Department of Physics, Budapest University of Technology and Economics and MTA-BME Lend\"ulet Magneto-optical Spectroscopy Research Group, 1111 Budapest, Hungary}
\affiliation{Hungarian Academy of Sciences, Premium Postdoctor Program, 1051 Budapest, Hungary}

\author{Y. Tokunaga}
\affiliation{RIKEN Center for Emergent Matter Science (CEMS), Wako, Saitama 351-0198, Japan}
\affiliation{Department of Advanced Materials Science, University of Tokyo, Kashiwa 277-8561, Japan}

\author{J. Viirok}
\affiliation{National Institute of Chemical Physics and Biophysics, Akadeemia tee 23, 12618, Tallinn, Estonia}

\author{L. Peedu}
\affiliation{National Institute of Chemical Physics and Biophysics, Akadeemia tee 23, 12618, Tallinn, Estonia}

\author{T. R\~o\~om}
\affiliation{National Institute of Chemical Physics and Biophysics, Akadeemia tee 23, 12618, Tallinn, Estonia}

\author{U. Nagel}
\affiliation{National Institute of Chemical Physics and Biophysics, Akadeemia tee 23, 12618, Tallinn, Estonia}

\author{Y. Taguchi}
\affiliation{RIKEN Center for Emergent Matter Science (CEMS), Wako, Saitama 351-0198, Japan}

\author{Y. Tokura}
\affiliation{RIKEN Center for Emergent Matter Science (CEMS), Wako, Saitama 351-0198, Japan}
\affiliation{Department of Applied Physics and Tokyo College, University of Tokyo, Hongo, Tokyo 113-8656, Japan}

\author{I. K\'ezsm\'arki}
\affiliation{Department of Physics, Budapest University of Technology and Economics and MTA-BME Lend\"ulet Magneto-optical Spectroscopy Research Group, 1111 Budapest, Hungary}
\affiliation{Experimental Physics 5, Center for Electronic Correlations and Magnetism, Institute of Physics,University of Augsburg, 86159 Augsburg, Germany}

\begin{abstract}
We have studied spin excitations in a single-domain crystal of  antiferromagnetic LiCoPO$_4$ by THz absorption spectroscopy. By analysing the selection rules and comparing the strengths of the absorption peaks in the different antiferromagnetic domains, we found electromagnons and magnetoelectric (ME) spin resonances besides conventional magnetic-dipole active spin-wave excitations.
Using the sum rule for the ME susceptibility we determined the contribution of the spin excitations to all the different off-diagonal elements of the static ME susceptibility tensor in zero as well as in finite magnetic fields.
We conclude that the ME spin resonances are responsible for the static ME response of the bulk, when the magnetic field is along the $x$ axis, and the symmetric part of the ME tensor with zero diagonal elements dominates over the antisymmetric components.
\end{abstract}

% PACS numbers
\maketitle

\section{Introduction}

The magnetoelectric (ME) effect is the cross induction of polarization and magnetization by magnetic and electric field, respectively, as described by the ME tensor forms, $P_\mu=\chi_{\mu\nu}H_\nu$ and $\mu_0M_\mu=\chi^{\rm T}_{\mu\nu}E_\nu$, where $P_\mu$ ($M_\mu$) and $H_\nu$ ($E_\nu$) are the $\mu,\nu = x,y,z$ components of the electric polarization (magnetic dipole moment) and the  magnetic (electric) field.
The ME effect is often associated with complex magnetic order parameters~\cite{Dubovik1990,Spaldin2008}, such as the ferrotoroidal~\cite{Aken2007,Rivera1994,Baum2013,Toledano2011,Toledano2015,Kezsmarki2011} and the ferroquadrupolar moments~\cite{Kato2017PRL}, or magnetically induced chirality~\cite{Bordacs2012}.
The ME effect provides a handle to manipulate these exotic spin orders and the corresponding magnetic domains even in the absence of spontaneous electric polarization or magnetization, thus, such ME materials have been expected as building blocks for novel data storage and memory devices~\cite{Borisov2005PRL,Shiratsuchi2018APL,Kocsis2018PRL}.
In some cases, such spin-multipolar orders have been revealed successfully by spherical neutron polarimetry~\cite{Baum2013,Babkevich2017PRL} and X-ray spectroscopy~\cite{Matteo2005}, and investigated indirectly using static ME measurements~\cite{Rivera1994,Ederer2009,Toledano2011,Toledano2015} and second-harmonic generation (SHG) microscopy~\cite{Fiebig2002a,Aken2007,Zimmermann2009}.

In this work, we exploit a different approach to assign ferrotoroidal and ferroquadrupolar orders, which is based on the measurement of the optical ME effect~\cite{Kezsmarki2011,Arima2008JPCM,Saito2008jpsj} using THz absorption spectroscopy.
In a ME medium counter-propagating light beams can experience different indices of refraction, exhibiting the optical directional anisotropy (ODA), as schematically shown in Fig.~\ref{lcpo01}(b).
This compelling phenomenon can be used to measure the dynamic ME response, also known as optical ME effect, e.g. in resonance with magnon modes at THz frequencies.
From the spectrum of the dynamic ME effect one can also determine the static ME coupling via the ME susceptibility sum rule~\cite{Szaller2014}.  
Furthermore, THz absorption spectroscopy can measure the ME domain population as it has been successfully utilized to distinguish between the two types of antiferromagnetic (AFM) domains of LiCoPO$_4$~\cite{Kocsis2018PRL} [see Fig.~\ref{lcpo01}(b)].
Here, in the case of LiCoPO$_4$, we demonstrate that the ODA can also be used to investigate the form and the spectral dependence of the ME susceptibility tensor and hence to identify different spin-multipolar orders responsible for the ME effect. 

\section{Advantage of optical over static ME experiments}

When the ME phase appears upon a second order phase transition from a high temperature centrosymmetric and paramagnetic phase, the ME domains ($\alpha$ and $\beta$) connected by the spatial
inversion and the time reversal symmetries have ME susceptibilities of opposite signs, $\hat{\chi}^{\alpha}=-\hat{\chi}^{\beta}$.
In the absence of electric ($E$) and magnetic ($H$) fields, a multi-domain state is often realized and the ME effect is canceled on the macroscopic scale.
When a material possesses ferroelectricity or ferromagnetism, the $P$ or $M$ domain with an order parameter parallel to the conjugate electric ($E^0$) or magnetic ($H^0$) field is selected, respectively.
However, when staggered electric and magnetic dipole orders or higher-order magnetic multipoles give rise to the ME effect, such direct control is not possible.
Instead, a single ME domain can be selected by the simultaneous application of $E^0$ and $H^0$ fields upon cooling a sample through the ordering temperature, which is often referred to as ME poling, while $E^0$ and $H^0$ as poling fields.

\begin{center}
    \begin{figure*}[t!]

    \includegraphics[width=17.0truecm]{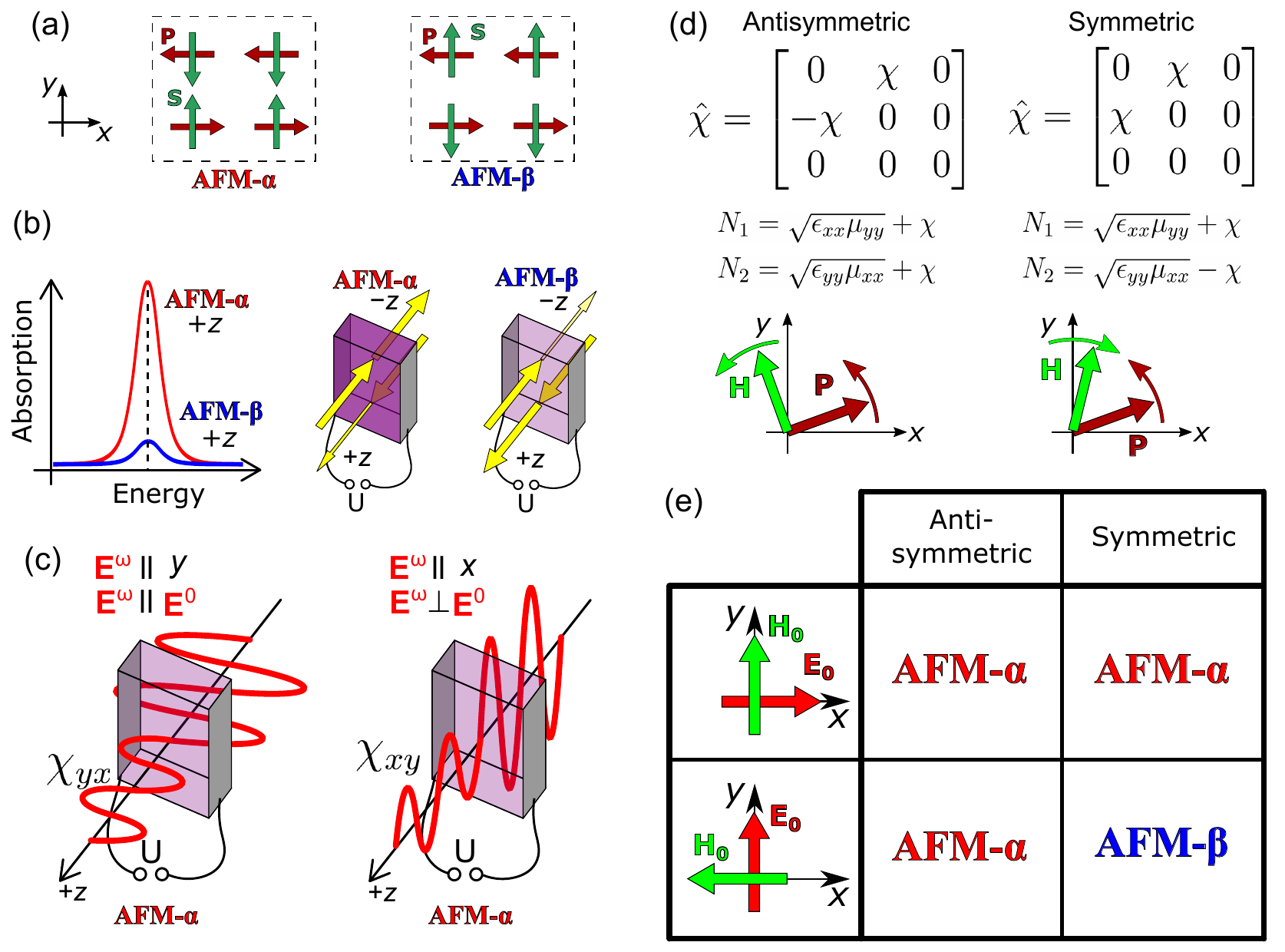}
    \caption{(Color online)
    (a)    
    The two AFM domains of LiCoPO$_4$; AFM-$\alpha$ and AFM-$\beta$ are characterized by ME coupling of opposite signs.
    (b)
    The dynamical magnetoelectric (ME) effect produces an absorption difference between light beams with the same polarizations propagating along the same directions ($+z$ or $-z$) in the two AFM domains.
    (c)
    Polarization of the probing light ($\mathbf{E}^\omega$) can be selected either parallel or perpendicular to the poling electric field ($\mathbf{E}^0$).
    Thus, in the dynamic ME measurement both elements of the ME susceptibility, $\chi_{xy}$ and $\chi_{yx}$, can be simultaneously measured.
    (d)
    Electric polarizations and indices of refractions [Eqs.~(\ref{eq:Nx}) and (\ref{eq:Ny})] resulting from symmetric or antisymmetric ME susceptibility tensors ($\mathbf{P}=\hat{\chi}\mathbf{H}$) rotate against or together with the magnetic field, respectively.
    (e)
    For both the symmetric and antisymmetric forms of $\hat{\chi}$, the poling-field combination ($+E^0_x$, $+H^0_y$) select the AFM-$\alpha$ domain.
    However, for the 90$^\circ$ rotated poling fields, poling with ($+E^0_y$, $-H^0_x$) selects domain AFM-$\alpha$ or AFM-$\beta$ in case of antisymmetric and symmetric $\hat{\chi}$, respectively.
    This is because with the former and later form of $\hat{\chi}$, the material couples to $E^0_x H^0_y \mp E^0_y H^0_x$, respectively.}
    \label{lcpo01}
    \end{figure*}
    \end{center}

In general, the ME susceptibility tensor is the sum of a traceless symmetric part ($\hat{\chi}^{\rm T}=\hat{\chi}$, quadrupolar part), an antisymmetric part ($\hat{\chi}^{\rm T}=-\hat{\chi}$, toroidal part), and diagonal (axion like) elements~\cite{Spaldin2008}.
For simplicity let us assume that the ME tensor has non-zero components only in the $xy$ plane.
The symmetric and antisymmetric ME susceptibilities, often associated with quadrupolar and toroidal spin orders, respectively, behave differently upon the rotation of external fields in the $xy$ plane, as shown in Fig.~\ref{lcpo01}(d).
If the magnetic field induced $P$ is described by an antisymmetric ME susceptibility, it rotates in the same sense as $H$ does, while it rotates in the opposite sense when generated by the symmetric, traceless ME susceptibility. As an example, if the ME susceptibility tensor is fully antisymmetric, the selected domain depends only on the cross-product of the poling $E^0$ and $H^0$ fields [see Fig.~\ref{lcpo01}(e)].
Thus, poling with orthogonal $E^0$ and $H^0$ fields of certain orientation selects the \textit{same} ME domain as poling with $E^0$ and $H^0$ fields mutually rotated by $90^\circ$. 
On the contrary, if the ME susceptibility tensor is symmetric and traceless as shown in Fig.~\ref{lcpo01}(d), $E^0$ and $H^0$ fields applied along the principal axes and the $90^\circ$ rotated poling fields prefer \textit{different} ME domains.

Static ME ($P$-$H$) measurements alone can usually provide a limited information about the form of the ME susceptibility tensor.
Since the same electric contacts are used to apply the $E^0$ poling field as well as to detect the magnetic field induced $P$, not all elements of the ME tensor can be measured in a single experimental configuration.
More specifically, in a different experimental configuration when the $E^0$ poling field is perpendicular to the magnetic field induced $P$, it is difficult to perform a reliable measurement.
In contrast, if the ME effect is detected optically via the ODA, the polarization of the probing light beam ($\mathbf{E}^\omega$) can be chosen independently of the poling field direction, either $\mathbf{E}^\omega\parallel\mathbf{E}^0$ or $\mathbf{E}^\omega\perp\mathbf{E}^0$, as illustrated in Fig.~\ref{lcpo01}(c).
Thus, both off-diagonal elements, $\chi_{xy}$ and $\chi_{yx}$, can be measured optically for a given ME domain selected by the poling.

\section{Static ME effect of L\lowercase{i}C\lowercase{o}PO$_4$}

The paramagnetic phase of LiCoPO$_4$ is described by a centrosymmetric and orthorhombic space group ($Pnma$), \textit{i.e.} this material does not have any spontaneous electric polarization.
The site symmetry of the magnetic Co$^{2+}$ ions allows local electric dipoles in the $xz$ plane, which are arranged in a staggered configuration on the four Co sites in the unit cell of this structure~\cite{Kocsis2018PRL}.
At $T_{\rm N}$=21.3\,K a four-sublattice N\'eel-type AFM order emerges with $S$=3/2 spins mainly co-aligned along the $y$ axis~\cite{Santoro1966,Vaknin2002}.
The two possible AFM domain states, $\alpha$ and $\beta$, which are also the two ME domains with opposite signs of $\hat{\chi}$, are illustrated in Fig.~\ref{lcpo01}(a).
In this compound, the magnetic order simultaneously breaks the inversion and the time-reversal symmetries, which allows finite $\chi_{xy}$ and $\chi_{yx}$ components of the ME tensor~\cite{Rivera1994}. Previously, the magnetically ordered state was identified as the first example of a ferrotoroidal order~\cite{Aken2007}, however, the form of the ME tensor, i.e.~the relative sign of $\chi_{xy}$ and $\chi_{yx}$ has remained an open question due to the experimental limitations discussed above~\cite{Zimmermann2014}.

We studied single crystal LiCoPO$_4$ samples that were grown by the optical floating zone method, similarly to the procedure described in Ref.~\onlinecite{SaintMartin2008}.
The ingots were aligned using a back-reflection Laue camera and cut into thin slabs with 1$\times$5$\times$5\,mm$^3$ dimensions.
Static magnetization measurements up to $H$=140\,kOe were done using a Physical Property Measurement System (PPMS, Quantum Design) equipped with a VSM option.
The magnetic-field induced polarization measurements were carried out in a PPMS using an electrometer (6517A, Keithley) in the charge ($Q$) measurement mode.

Following the application of orthogonal poling fields ($\mathbf{E}^0\parallel{y}$, $\mathbf{H}^0\parallel{x}$) and ($\mathbf{E}^0\parallel{x}$, $\mathbf{H}^0\parallel{y}$) we measured the ME susceptibility as shown in Fig.~\ref{lcpo02}(c) and \ref{lcpo02}(d), respectively.
The experimental configurations are illustrated in Fig.~\ref{lcpo02}(a) and \ref{lcpo02}(b), respectively.
In both orientations, the measurement was carried out in all four different poling configurations, namely, with selectively reversed signs of the $E^0$ and $H^0$ fields.
In the ordered phase $\mathbf{E}^0$ was switched off, and the displacement-current measurements were done in sweeping $H$ field between $\pm$1\,kOe for five times.
The magnitudes of the measured ME susceptibilies at $T$=2\,K are $\vert\chi_{xy}\vert/c$=15\,ps/m and $\vert\chi_{yx}\vert/c$=32\,ps/m, and agree well with those previously reported in the literature~\cite{Rivera1994}.
Poling $E^0$ and $H^0$ fields of the same sign select one ME domain, while poling fields of opposite signs select the other ME domain~\cite{Kocsis2018PRL}.
However, due to the experimental limitations inherent to the static ME experiments as described above, only absolute value of one of the two finite off-diagonal components of $\hat{\chi}$ can be measured for a given orientation of the poling fields.
In contrast, if the ME effect is investigated optically one can determine both, $\chi_{xy}$ and $\chi_{yx}$, for each poled state, as will be discussed in details in the following.
This only requires the rotation of the light polarization by $90^\circ$ in the plane of the poling fields.

\section{Optical determination of the ME susceptibility tensor}

Spin excitations with ME character, the ME resonances~\cite{Kezsmarki2011,Takahashi2012}, can be simultaneously excited by the electric ($\mathbf{E}^\omega$) and magnetic ($\mathbf{H}^\omega$) components of light and, therefore, can be exploited to probe the elements of the dynamic ME susceptibility tensor.
Such spin resonances can show strong absorption difference for the respective ME domains, due to the opposite signs of the ME susceptibility in the two domains, $\alpha$ and $\beta$.
When light propagates in such material, the oscillating magnetization in both the $\alpha$ and $\beta$ domains fluctuate in phase with $\mathbf{H}^\omega$, while the corresponding magnetically induced polarizations, in the two domains oscillate in anti-phase with respect to each other.
As a result, the index of refraction for light propagation along the $+z$ axis of the crystal is different for the two domains~\cite{Wooten1972,KocsisPHD}:
\begin{align}
N_1^{\alpha/\beta}(\omega) = \sqrt{\epsilon_{xx}(\omega)\mu_{yy}(\omega)}+\chi^{\alpha/\beta}_{xy}(\omega), \mathrm{~for~}E^\omega_{x},H^\omega_{y} \label{eq:Nx}\\ 
N_2^{\alpha/\beta}(\omega) = \sqrt{\epsilon_{yy}(\omega)\mu_{xx}(\omega)}-\chi^{\alpha/\beta}_{yx}(\omega), \mathrm{~for~}E^\omega_{y},H^\omega_{x}\label{eq:Ny}
\end{align}
where  the $\epsilon_{\nu\nu}$ and $\mu_{\nu\nu}$, ($\nu=x,y$) are elements of the dielectric permittivity and magnetic permeability tensors, respectively, as well as $E^\omega_{\nu}$ , $H^\omega_{\nu}$ denotes the $\nu$ component of the oscillating electric and magnetic fields.
The light absorption is different for the two ME (AFM) domains as the ME susceptibility has opposite sign for them, $\hat{\chi}^{\alpha}=-\hat{\chi}^{\beta}$. 
We note that the reversal of the light propagation direction from $+z$ to $-z$ is equivalent to the exchange of the ME domains (Fig.~\ref{lcpo01}a), thus the ODA also has opposite sign for the two ME domains.
The sign difference between Eqs.~(\ref{eq:Nx}) and (\ref{eq:Ny}) is related to the rotation of the light polarization.
From Eqs.~(\ref{eq:Nx}) and (\ref{eq:Ny}), it follows that in materials with an antisymmetric ME effect the differences in the refractive indices of the two AFM domains, $\Delta{N}_1=(N_1^\alpha-N_1^\beta)/2$ and $\Delta{N}_2=(N_2^\alpha-N_2^\beta)/2$, are the same for the two orthogonal light polarizations, $\Delta{N}_1=\Delta{N}_2=\chi_{xy}$.
On the other hand, for systems with symmetric ME susceptibility tensor, $\chi_{xy}=\chi_{yx}$ the differences in the refractive indices of the two domains changes sign, $\Delta{N}_1=-\Delta{N}_2$, upon the rotation of light polarization by $90^\circ$ .
We note here that, such changes in the ODA have to be probed on a single excitation, as the sign of the optical ME susceptibility is specific to the different excited states.

\section{Magnons, electromagnons, and ME resonances in L\lowercase{i}C\lowercase{o}PO$_4$}

\begin{center}
    \begin{figure*}[t!]

    \includegraphics[width=15.0truecm]{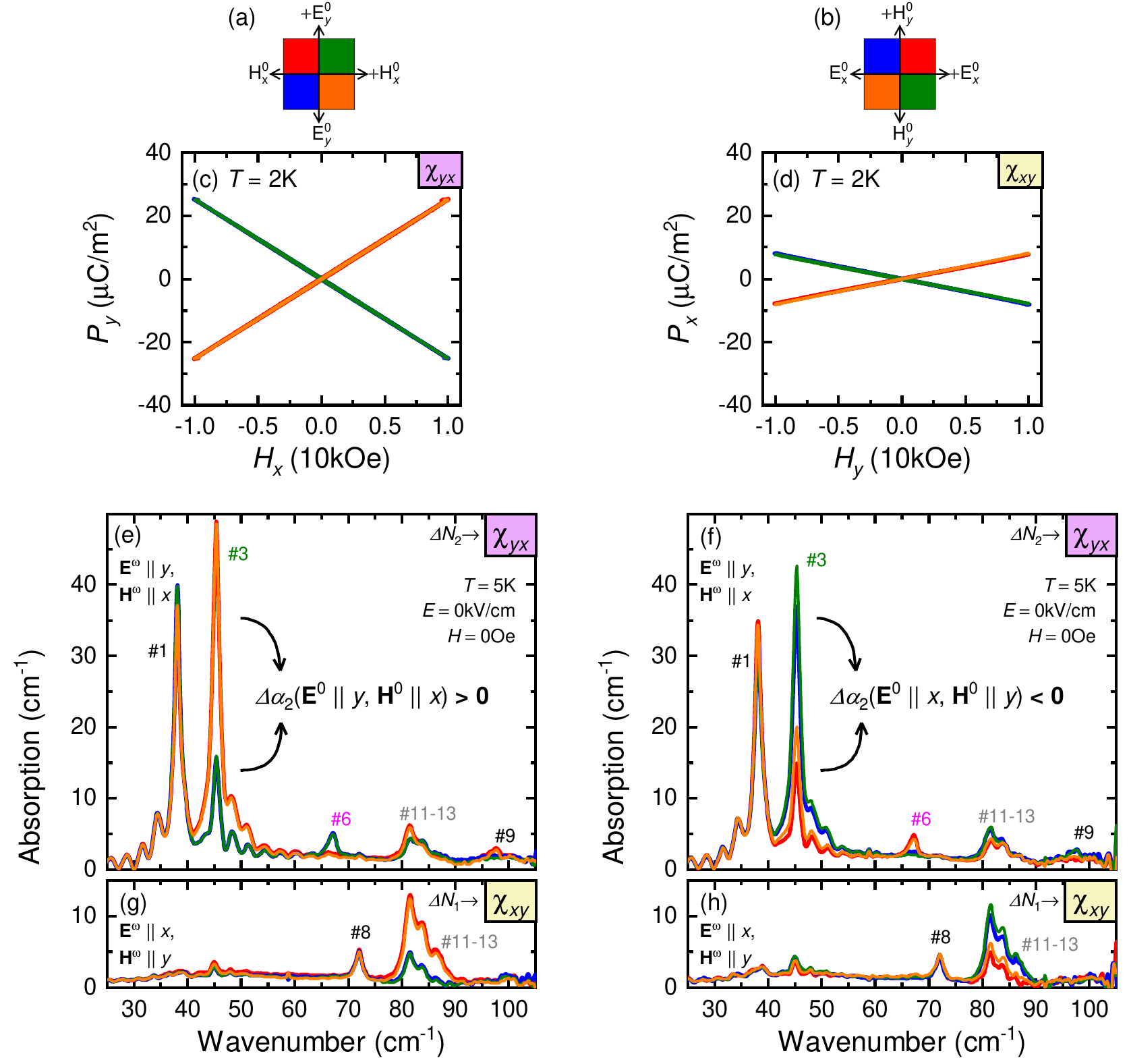}
    \caption{(Color online) Remanent static and dynamic ME effects in LiCoPO$_4$.
    (a,b) Experimental configuration of the $\mathbf{E}^0$ and $\mathbf{H}^0$ poling fields. The color code of panel (b) corresponds to the simultaneous 90 degree rotation of the fields of panel (a).
    (c,d) Poling field dependence of the $P$-$H$ curves at $T$=5\,K for (c) $\mathbf{E}^0\parallel{y}$ and $\mathbf{H}^0\parallel{x}$ and (d) $\mathbf{E}^0\parallel{x}$ and $\mathbf{H}^0\parallel{y}$ poling field configurations.
    Color code of panels (c) and (d) is shown in panels (a) and (b), respectively.
    The slopes of the $P$-$H$ curves correspond to $\chi_{yx}$ and $\chi_{xy}$ in panels (c) and (d), respectively.
    The sign of the ME effect depends on the relative signs of the poling fields, note the complete overlap of the red and orange, as well as the blue and green curves.
    (e-h)
    Optical absorption spectra measured at $T$=5\,K (e,g) with poling configurations indicated in panel (a) and (f,h) with poling configurations shown in panel (b).
    Spectra in panels (e,f) were measured using linearly polarized light with $E^\omega_{y}$, $H^\omega_{x}$, while in panels (g,h) with  $E^\omega_{x}$, $H^\omega_{y}$.
    The spectra of the dynamic ME coefficients can be calculated as the absorption difference of the different domains, according to Eqs.~(\ref{eq:Nx}) and (\ref{eq:Ny}).
    The $\Delta\alpha_2$ ODA in panels (e) and (f) as well as in (g) and (h) change sign for the rotation of the poling $E^0$ and $H^0$ poling fields, according to Eqs.~(\ref{eq:DN2_1}) and (\ref{eq:DN2_2}), respectively.}
    \label{lcpo02}
    \end{figure*}
    \end{center}

Optical absorption spectra of LiCoPO$_4$ were measured at the National Institute of Chemical Physics and Biophysics, Tallinn using a Martin-Puplett interferometer combined with a superconducting magnet, applying  magnetic fields up to $H$=170\,kOe. The relative absorption spectra recorded at $T$=5\,K, using linearly polarized light with $\mathbf{E}^\omega\parallel{y}$ and $x$ are shown in Figs.~\ref{lcpo02}(e,f) and \ref{lcpo02}(g,h), respectively.
The sample was cooled to a ME single-domain state in $E^0$=1\,kV/cm and $H^0$=1\,kOe poling fields, respectively applied along the ${y}$ and ${x}$ axes.
The low-temperature absorption measurements were carried out after switching off the poling fields.
The relative absorption spectra were obtained by subtracting a reference spectrum taken in the paramagnetic phase, at $T$=30\,K.
Thus, the low-temperature spectral features are related to excitations emerging in the magnetically ordered state:
Two strong ($\#$1 and $\#$3) and several weaker ($\#$6, $\#$8, $\#$9, and $\#$11-13) resonances appear in the AFM phase.

The poling-field dependent resonances, $\#$1, $\#$3, $\#$6, $\#$9 and $\#$11-13 are ME resonances since the ME response has the opposite sign in the $\alpha$ and $\beta$ domains. For the same signs of the poling fields, $(+E_y^0,+H_x^0)$ and $(-E_y^0,-H_x^0)$, modes $\#$3, $\#$9 and $\#$11-13 have large absorptions, while for opposite signs of the poling fields, $(-E_y^0,+H_x^0)$ and $(+E_y^0,-H_x^0)$, the same modes show lower absorption.
As the magnitude of the absorption difference for the ME domains is the highest for resonance $\#$3, in the following we will focus on this mode.
It appears when light polarization is $\mathbf{E}^\omega\parallel{y}$ and $\mathbf{H}^\omega\parallel{x}$, thus, according to Eq.~(\ref{eq:Ny}), it probes $\chi_{yx}(\omega)$. For static poling fields $\mathbf{E}^0\parallel{y}$ and $\mathbf{H}^0\parallel{x}$ the difference of the absorption coefficients $\Delta\alpha=\frac{2\omega}{c}Im(\Delta{N})$,
\begin{equation}
\Delta{\alpha}_2=[\alpha_2(+E^0_y,-H^0_x)-\alpha_2(-E^0_y,-H^0_x)]/2 >0,\label{eq:DN2_1}
\end{equation}
$\Delta{\alpha}_2$ is positive, as clear from Fig.~\ref{lcpo02}(e) while for poling fields rotated by 90$^\circ$ to $\mathbf{E}^0\parallel{x}$ and $\mathbf{H}^0\parallel{y}$,
\begin{equation}
\Delta{\alpha}_2=[\alpha_2(+E^0_x,+H^0_y)-\alpha_2(-E^0_x,+H^0_y)]/2 <0,\label{eq:DN2_2}
\end{equation}
$\Delta{\alpha}_2$ is negative, as seen in Fig.~\ref{lcpo02}(f).
From this we can conclude that if poling with ($+E^0_y$,$-H^0_x$) and ($-E^0_y$,$-H^0_x$) has selected domains $\alpha$ and $\beta$, respectively, then poling with ($+E^0_x$,$+H^0_y$) and ($-E^0_x$,$+H^0_y$) must have selected domains $\beta$ and $\alpha$.
It means that rotation of the poling fields by 90$^\circ$ results in the selection of a different ME domain, as illustrated in Fig.~\ref{lcpo01}(e).
Correspondingly, the symmetric traceless part of the ME tensor governs the poling, \textit{i.e.} the ME order parameter couples more efficiently to the symmetric product of the poling fields, ($E^0_xH^0_y+E^0_yH^0_x$), as illustrated in Fig.~\ref{lcpo01}(d,e).
Since neither component of the ME susceptibility changes sign~\cite{Rivera1994}, we concluded that the symmetric part dominates the low temperature ME tensor.
As a result, the magnitudes of the traceless symmetric and antisymmetric components of the static ME susceptibility are estimated to be $\chi^{\rm symm}/c$=23.5\,ps/m and $\chi^{\rm antisymm}/c$=8.5\,ps/m, respectively, based on the static measurements.
The symmetric part is about 2.8 times as large as the antisymmetric.

\begin{center}
    \begin{figure*}[h]

    \includegraphics[width=17.0truecm]{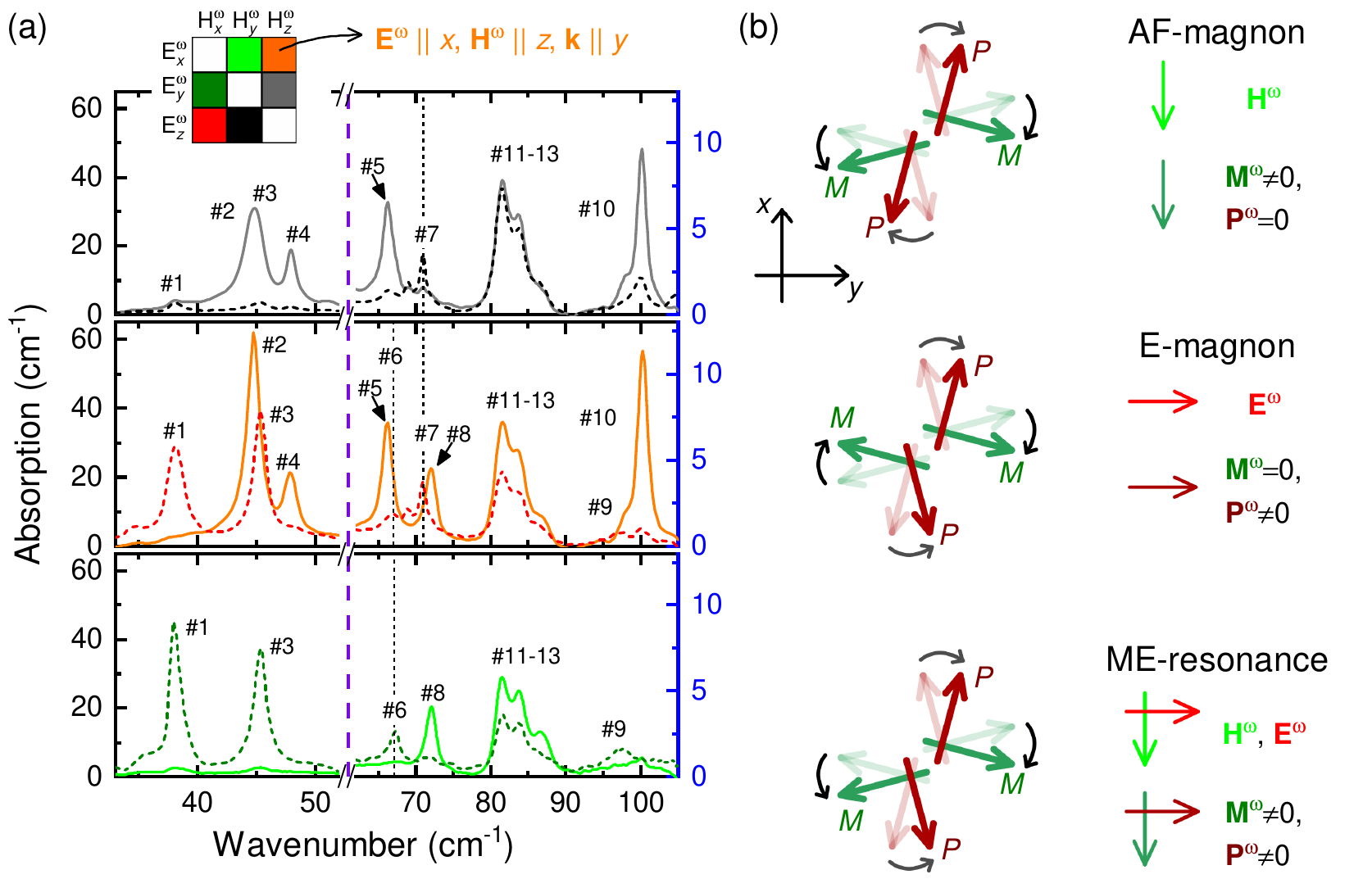}
    \caption{(Color online) (a) Optical absorption spectra, measured in six different configurations of the $\mathbf{E}^\omega$ and $\mathbf{H}^\omega$ of the light at $T$=5\,K, revealing the selection rules of the magnetic excitations. The inset indicates the color code of the different light polarizations, along with an example, where $\mathbf{E}^\omega\parallel{x}$, $\mathbf{H}^\omega\parallel{z}$, and $\mathbf{k}\parallel{y}$. Note the difference in the vertical scales corresponding to the spectral regions below and above the vertical dashed line.
    (b) Schematic illustration of an antiferromagnetic resonance (AF-magnon), an electromagnon (E-magnon), and a ME-resonance. Respectively, these modes are excited by only the magnetic, the electric, and simultaneously by both components of the electromagnetic radiation. For the sake of simplicity, we illustrate the excitations with one representative pair of spin ($M$) of the magnetic unit cell along with the respective local polarization ($P$). The dynamic nature of the excitations is captured by curved arrows, which represent in-phase or anti-phase oscillations of the $M$ and $P$. The net dynamic magnetization and polarization of the unit cell are labeled as $\mathbf{M}^\omega$ and $\mathbf{P}^\omega$, respectively.}
    \label{lcpo03}
    \end{figure*}
    \end{center}

The selection rules were further studied by recoding the absorption spectra with light polarized along all the principal axes. The results of this systematic study are summarized in Fig.~\ref{lcpo03}(a) and in Table~\ref{tab:tableSelRule}.
In total, thirteen resonances are observed, one more than expected from the multi-boson spin-wave theory of a four sub-lattice AFM with $S$=3/2 spins~\cite{Kocsis2018PRL}.
Since the structural symmetry is preserved, no new phonon modes are expected in the magnetically ordered phase, thus the origin of the extra mode is unclear.
Figure \ref{lcpo03}(b) schematically illustrates the character of the different excitations.
In case of usual zone-center magnon modes of antiferromagnets, there is a finite $\mathbf{H}^\omega$ induced magnetization in each unit cell, as the different magnetic sublattices oscillate in phase.
There may be a dynamic electric polarization associated with the presence of individual spins, but these local polarizations oscillate out of phase and averages to zero over the unit cell.
Thus these modes only couple to uniform $\mathbf{H}^\omega$ but not $\mathbf{E}^\omega$.
In contrast for an electromagnon (E-magnon), responding only to $\mathbf{E}^\omega$, there is a finite electric polarization of the unit cell induced by the spin dynamics, but the dynamic magnetization is canceled due to the out-of-phase oscillation of the different sublattices.
In case of ME-resonances, both the dynamic magnetization and polarization of the unit cell are finite, thus these modes can be excited by $\mathbf{H}^\omega$ as well as $\mathbf{E}^\omega$.
Modes $\#$2 and $\#$4 are usual magnon modes which are excited only by the oscillating magnetic field of light~\cite{Kocsis2018PRL}. 
Modes $\#$1, $\#$3, $\#$6 and $\#$9 are ME resonances, as they exhibit ODA and are excited by both the $H^\omega_{x}$ and the $E^\omega_{y}$ components of light.
Mode $\#$8 is an E-magnon, as it is excited only by $E^\omega_{x}$.
Modes $\#$11-13 do not show simple selection rules but appear simultaneously for any polarization of the light and they exhibit ODA for $H^\omega_{y}$, thus they are ME resonances.
In summary, four magnon modes are excited by $H^\omega_z$ ($\#2$, $\#4$, $\#5$, and $\#10$), four ME-resonances are excited by $H^\omega_x$ as well as $E^\omega_y$ ($\#1$, $\#3$, $\#6$, and $\#9$), and two  E-magnons are excited by $E^\omega_z$ and $E^\omega_x$ ($\#7$ and $\#8$, respectively).

\section{Field dependence of the spin-wave excitations}

\begin{center}
    \begin{figure*}[t!]

    \includegraphics[width=17.0truecm]{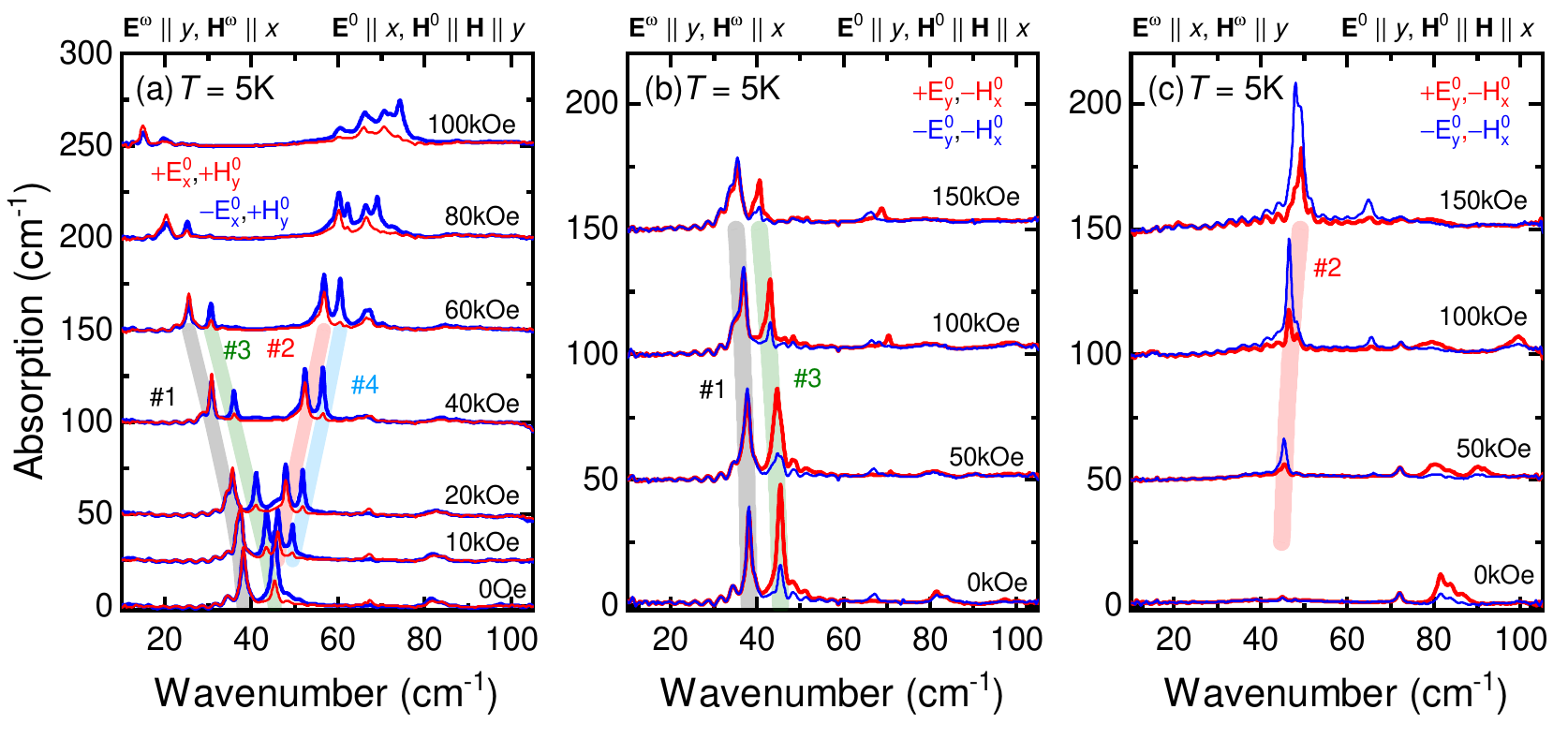}
    \caption{(Color online) Magnetic field dependence of the optical absorption spectra measured at $T$=5\,K for different combinations of the poling fields. The reference spectra, taken in the paramagnetic phase ($T$=30\,K), were subtracted from the spectra measured at $T$=5\,K. The poling fields were applied in the ($\mathbf{E}^0\parallel{x}$, $\mathbf{H}^0\parallel{y}$) and ($\mathbf{E}^0\parallel{y}$, $\mathbf{H}^0\parallel{x}$) configurations in panels (a) and (b,c), respectively; the static magnetic field applied after poling was $\mathbf{H}\parallel\mathbf{H}^0$. The $\mathbf{E}^\omega$ polarization of the electromagnetic radiation was set along the $y$ and $x$ axes, respectively for panels (a,b) and (c). For the purpose of clarity, each spectrum is shifted in proportion to the applied field by 25\,cm$^{-1}$/10\,kOe and 10\,cm$^{-1}$/10\,kOe for (a) and (b,c), respectively.}
    \label{lcpo04}
    \end{figure*}
    \end{center}

The characters and the frequencies of the spin-wave excitations together with their dynamic ME effect are further investigated using magnetic field dependent absorption measurements shown in Fig.~\ref{lcpo04}.
The magnetic field dependence of the absorption spectra, measured at $T$=5\,K, is presented in Figs.~\ref{lcpo04}(a,b) and \ref{lcpo04}(c), with $\mathbf{E}^\omega\parallel{y}$ and $\mathbf{E}^\omega\parallel{x}$, respectively.
The reference signal was again recorded in the paramagnetic phases at $T$=30\,K and subtracted from the $T$=5\,K spectra.
The external magnetic field was applied in the same direction as the magnetic field used for poling, $H\parallel{H}^0$.
Since the external magnetic field does not change the magnetic phase of the sample at low temperature~\cite{Kharchenko2010}, the domain state selected by the poling is preserved during the field dependent measurements.
The poling fields $E^0_x$=1\,kV/cm and $H^0_y$=100\,kOe were applied in Fig.~\ref{lcpo04}(a).
In this case, the two modes observed in zero field split into four distinct excitations.
Modes $\#$1 and $\#$3 shift to lower energies in proportion to the magnetic field, while modes $\#$2 and $\#$4 shift to higher energies.
In experiments corresponding to Figs.~\ref{lcpo04}(b) and \ref{lcpo04}(c), the poling fields of the same magnitude were applied in the perpendicular configuration, \textit{i.e.} $\mathbf{E}^0\parallel{y}$ and $\mathbf{H}^0\parallel{x}$.
In Fig.~\ref{lcpo04}(b), only modes $\#$1 and $\#$3 are observed.
With increasing magnetic field, they soften weakly and lose oscillator strength, while in Fig.~\ref{lcpo04}(c) only mode $\#$2 appears and slightly shifts to higher energies.
In zero magnetic field, mode $\#$2 is a usual magnon with no ODA, however, due to hybridization to modes $\#$1 and $\#$3 it also shows considerable directional effect in finite magnetic fields.

According to the sum rule established in Ref.~\onlinecite{Szaller2014}, excitations with ME character contribute to the static ME effect.
\begin{equation}
\chi_{ij}(0)=\frac{c}{2\pi}\int_0^\infty\frac{\Delta\alpha(\omega)}{\omega^2}d\omega,
\end{equation}
where $\Delta\alpha(\omega)$ is the absorption difference caused by the ODA for light polarization $\mathbf{E}^\omega_i$ and $\mathbf{H}^\omega_j$.
If the optical transitions are well separated in energy, it is possible to estimate the weight of each excitation to the static ME effect by limiting the integrate around the excitation.
The contributions of the respective resonances for domain $\alpha$ are denoted as $\tilde{\chi}^\alpha$.

\begin{center}
    \begin{figure*}[t!]

    \includegraphics[width=17.0truecm]{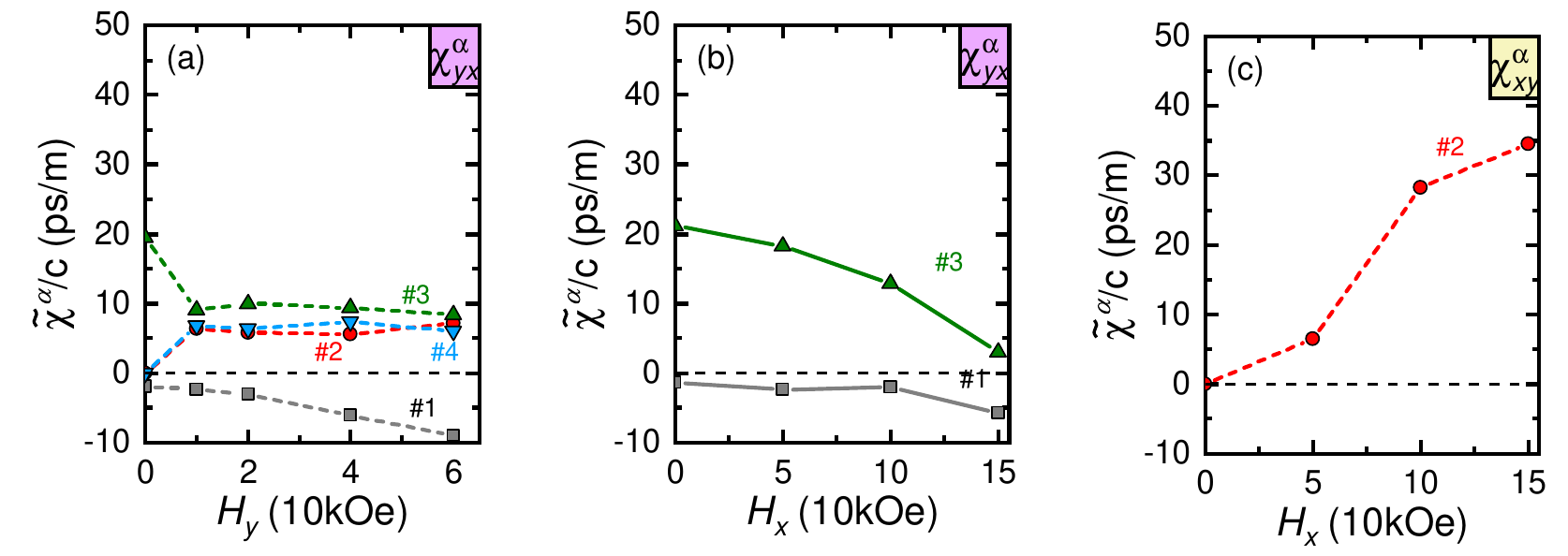}
    \caption{(Color online) 
    	Individual contributions of the different ME spin resonances to the static ME effect, calculated using the ME sum rule. 
		(a,b) Field dependence of $\chi_{yx}$ for the domain $\alpha$, with $\mathbf{H}\parallel{y}$ and $\mathbf{H}\parallel{x}$, respectively.
     	(c) Field dependence of $\chi_{xy}$ for the domain $\alpha$, with $\mathbf{H}\parallel{x}$. The color coding and the numbering of the curves correspond to the labeling of the modes in Fig.~\ref{lcpo04}. For $\chi^\alpha_{yx}$ and $\chi^\alpha_{xy}$ the sum of the different contributions and their field and temperature dependence are shown in Fig.~\ref{lcpo06} in comparison with the data from the static measurements.}
    \label{lcpo05}
    \end{figure*}
    \end{center}

The individual contributions of modes $\#$1 - $\#$4 to the static ME susceptibility estimated from the data presented in Fig.~\ref{lcpo04}(a) for $\mathbf{H}^0\parallel{y}$ are shown in Fig.~\ref{lcpo05}(a).
In this case the polarization of light $\mathbf{E}^\omega$ is perpendicular to the corresponding $\mathbf{E}^0$ poling field and likewise, $\mathbf{H}^\omega\perp\mathbf{H}^0$, which corresponds to the transverse ME susceptibility in the static limit.
The usual magnons, modes $\#$2 and $\#$4, which are forbidden in zero magnetic field for the polarization $E^\omega_x$ and $H^\omega_y$, gain optical weight as well as finite contribution to the $\chi^\alpha_{yx}$ ME susceptibility in finite fields (for further details see Fig.~S3).
For fields larger than $H$=10\,kOe, modes $\#$2, $\#$3, and $\#$4 have roughly equal and field-independent contribution to the static ME effect.
In contrast to the other excitations, mode $\#$1 has a negative and increasing contribution in larger $H$ field.
%The increase in the static ME effect with increasing field is mainly due to the softening of mode $\#$1, as it has a dominant contribution to the static ME effect due to the $1/\omega^2$ term in the ME sum rule (see Fig.~S3).

Figures~\ref{lcpo05}(b) and \ref{lcpo05}(c) show the field dependence of the $\tilde{\chi}^\alpha_{yx}$ and $\tilde{\chi}^\alpha_{xy}$ ME susceptibilities for $\mathbf{H}\parallel{x}$ field in the form of individual contributions from the different modes calculated using the ME sum rule on the data in Figs.~\ref{lcpo04}(b) and \ref{lcpo04}(c), respectively.
Modes $\#$1 and $\#$3 give contribution only to $\chi^\alpha_{yx}$, while $\chi^\alpha_{xy}$ is dominated by mode $\#$2.
Note that the contributions to $\chi^\alpha_{yx}$ from modes $\#$1 and $\#$3 have opposite signs for any direction of the magnetic field.

\begin{center}
    \begin{figure*}[t!]

    \includegraphics[width=17.0truecm]{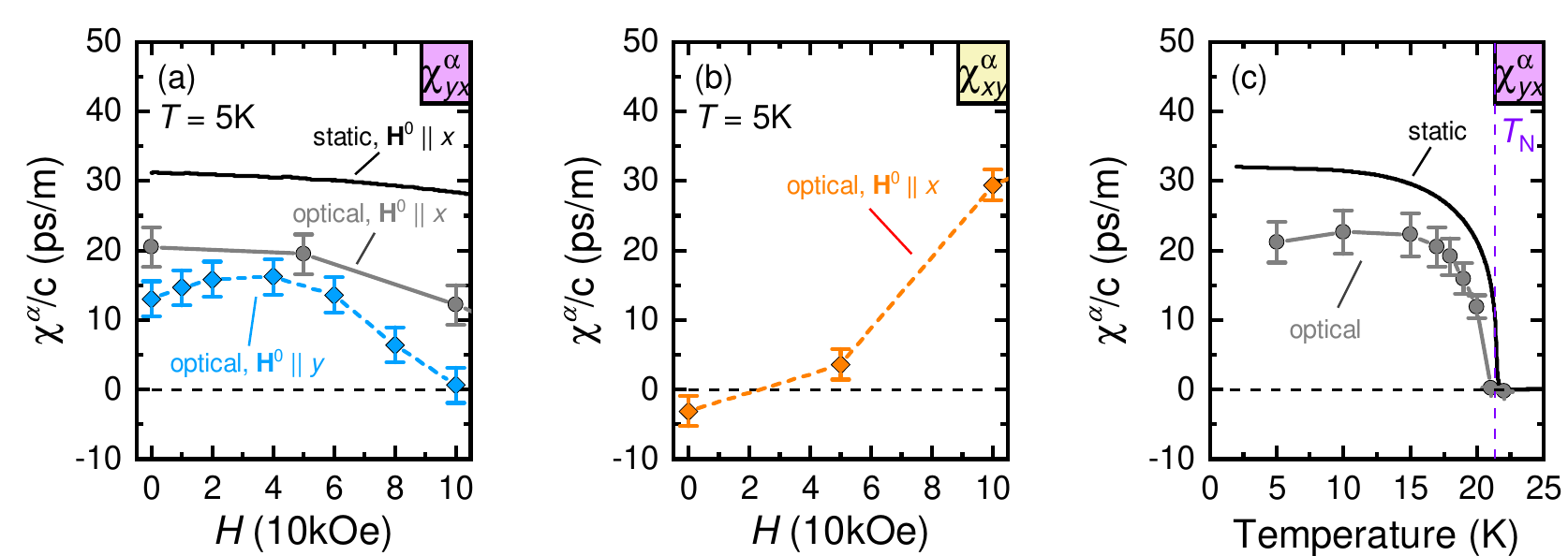}
    \caption{(Color online) Magnetic field (a,b) and temperature (c) dependence of the static and optical ME susceptibilities. The optical data was calculated using the ME sum rule over the complete spectral window. In panel (a), the grey and blue curves, corresponding to the field dependence of $\chi^\alpha_{yx}$ for $\mathbf{H}\parallel{x}$ and $\mathbf{H}\parallel{y}$, respectively, should coincide at $H=0$. Their slight deviation may come \textit{e.g.} from the imperfectness of the ME single-domain state in one of the measurements.
Panel (b) shows the magnetic field dependence of $\chi^\alpha_{xy}$ in applied field $\mathbf{H}\parallel\mathbf{H}^0\parallel{x}$.    
    In panel (c), the optical spectroscopy  measurements were carried out in the absence of $H$ field, while static ME susceptibility was taken in warming runs with $H$=10\,kOe field. The corresponding, temperature dependent optical absorption spectra can be found in Fig.~S2.
     The error bars were estimated using the average of the ME susceptibilities between $\pm{H}$ fields.}
    \label{lcpo06}
    \end{figure*}
    \end{center}

In Fig.~\ref{lcpo06} the zero-field static ME susceptibility obtained from magneto-current measurements and the sum of the contributions of the studied optical modes are compared.
Such a value obtained in via spectroscopy, $\chi_{yx}/c$=+20.5$\pm$2.9\,ps/m, is in a rather good agreement with the static value $\vert\chi_{yx}\vert/c$=32\,ps/m.
In this case the observed ME resonances explain well the bulk of the static ME response, i.e. the  polarization is mainly induced by the collective motion of spin-waves.
On the other hand, $\chi_{xy}/c$=$-$3.1$\pm$2.2\,ps/m, which is deduced from the optical experiments, is much smaller than the ME susceptibility $\vert\chi_{xy}\vert/c$=15\,ps/m measured in the static limit.
The most likely explanation of this discrepancy is that additional ME mode(s) or ME electronic excitations lie outside of the limited frequency range of our THz absorption measurement.

\section{Conclusions}

The antiferromagnetic LiCoPO$_4$ has two possible antiferromagnetic domain states with opposite signs of the ME coupling, which can be selected by the simultaneous application of $\mathbf{E}^0$ and $\mathbf{H}^0$ poling fields orthogonal to each other.
When selecting one of the domains by ME poling, the material shows optical directional anisotropy without any external fields, where the more transparent and absorbing directions are inter-changed for the two domains.
Using straightforward measurements of THz optical absorption after applying different ME poling configurations, we have found that the relative sign of the two allowed static ME susceptibility terms, $\chi_{yx}$ and $\chi_{xy}$, is the same.
According to our findings the magnetic order promotes a cross-coupling between electric and magnetic degrees of freedoms, which is described by a tensor with symmetric (quadrupolar) component larger than the antisymmetric (torroidal) component.
On the basis of the THz spectrum of ME effect we could explain the static ME effect $\chi_{yx}$ using a sum rule, while for $\chi_{xy}$ the relavant ME modes are out of the range of our experiment.
This optical method can be utilized to determine all off-diagonal elements of the ME susceptibility in a wide range of ME materials.

\begin{acknowledgments}
The authors are grateful to J.~Romh\'anyi and K.~Penc for discussion and for the technical assistance provided by A.~Kikkawa and M.~Kriener. V. Kocsis was supported by RIKEN Incentive  Research Project FY2016. This project was supported by institutional research funding IUT23-3 of the Estonian Ministry of Education and Research, by the European Regional Development Fund project TK134, by the bilateral program of the Estonian and Hungarian Academies of Sciences under the Contract No. NKM-47/2018, by the BME-Nanonotechnology and Materials Science FIKP grant of EMMI (BME FIKP-NAT), and by the Deutsche Forschungsgemeinschaft  (DFG)  via  the  
Transregional  Research Collaboration TRR 80: From Electronic
Correlations to Functionality (Augsburg-Munich-Stuttgart).
\end{acknowledgments}

\cleardoublepage
\pagestyle{empty}

%\bibliographystyle{apsrev4-1}
%\bibliography{Multiferroics}

    %***********************************************************************************************************************************************
    %FIG #1
    %***********************************************************************************************************************************************

    %***********************************************************************************************************************************************
    %FIG #2
    %***********************************************************************************************************************************************

    %***********************************************************************************************************************************************
    %FIG #3
    %***********************************************************************************************************************************************

\begin{table}[t]
\begin{tabular}{||p{15mm}||p{35mm}|p{25mm}|p{30mm}||}
\hline
\hline
Mode & Excitation & Remanent ODA & Class\\
\hline
\hline
$\#1$ & $\mathbf{H}^\omega\parallel{x}$, $\mathbf{E}^\omega\parallel{y}$ & $+$ODA (small) & ME-resonance\\
\hline
$\#2$ & $\mathbf{H}^\omega\parallel{z}$ & no ODA & magnon\\
\hline
$\#3$ & $\mathbf{H}^\omega\parallel{x}$, $\mathbf{E}^\omega\parallel{y}$ & $-$ODA (large) & ME-resonance\\
\hline
$\#4$ & $\mathbf{H}^\omega\parallel{z}$ & no ODA & magnon\\
\hline
$\#5$ & $\mathbf{H}^\omega\parallel{z}$ & no ODA & magnon\\
\hline
$\#6$ & $\mathbf{H}^\omega\parallel{x}$ & $+$ODA (100$\%$) & ME-resonance\\
\hline
$\#7$ & $\mathbf{E}^\omega\parallel{z}$ & no ODA & electromagnon\\
\hline
$\#8$ & $\mathbf{E}^\omega\parallel{x}$ & no ODA & electromagnon\\
\hline
$\#9$ & $\mathbf{H}^\omega\parallel{x}$ & $-$ODA (100$\%$) & ME-resonance\\
\hline
$\#10$ & $\mathbf{H}^\omega\parallel{z}$ & no ODA & magnon\\
\hline
$\#$11-13 & present in\newline each polarization & ODA & character cannot be determined\\
\hline
\hline
\end{tabular}
\caption{Summary on the magnetic excitations in terms of exciting fields, optical directional anisotropy (ODA) and classification of the resonances. The table is based on the zero-field measurements shown in Fig.~\ref{lcpo03}}\label{tab:tableSelRule} 
\end{table}

    %***********************************************************************************************************************************************
    %FIG #4
    %***********************************************************************************************************************************************

    %***********************************************************************************************************************************************
    %FIG #5
    %***********************************************************************************************************************************************

    %***********************************************************************************************************************************************
    %FIG #6
    %***********************************************************************************************************************************************

\cleardoublepage

\newpage
\newpage

\renewcommand{\thefigure}{S\arabic{figure}}
\renewcommand{\theequation}{S\arabic{equation}}
\renewcommand{\thetable}{S\arabic{table}}
\setcounter{figure}{0}
\cleardoublepage

\begin{center}
\textbf{Supplementary Material}
\end{center}

    %***********************************************************************************************************************************************
    %SFIG #1
    \begin{center}
    \begin{figure}[h]

    \includegraphics[width=8.0truecm]{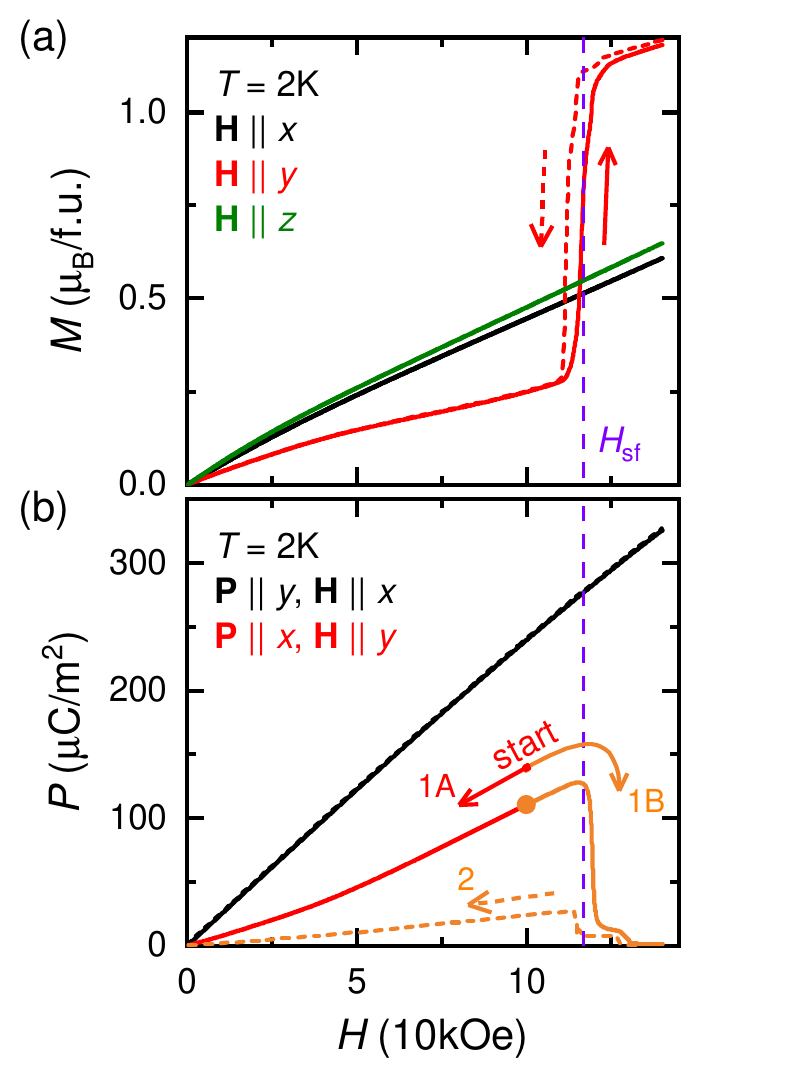}
    \caption{(Color online) Magnetic field dependence of the (a) magnetization and (b) polarization measured along the main crystallographic axes at $T$=2\,K. The spin-flop phase transition at $H_{1,+}$=116\,kOe is indicated by vertical dashed line. After cooling in ($+E^0$,$+H^0$) poling fields the $x$ component of the polarization was measured starting from $H_y^0$=$+$100\,kOe in field decreasing or increasing loops, respectively labeled by roman and arabic numbers.}
    \label{lcpoS01}
    \end{figure}
    \end{center}
    %***********************************************************************************************************************************************

    %***********************************************************************************************************************************************
    %SFIG #2
    \begin{center}
    \begin{figure*}[h]

    \includegraphics[width=15.0truecm]{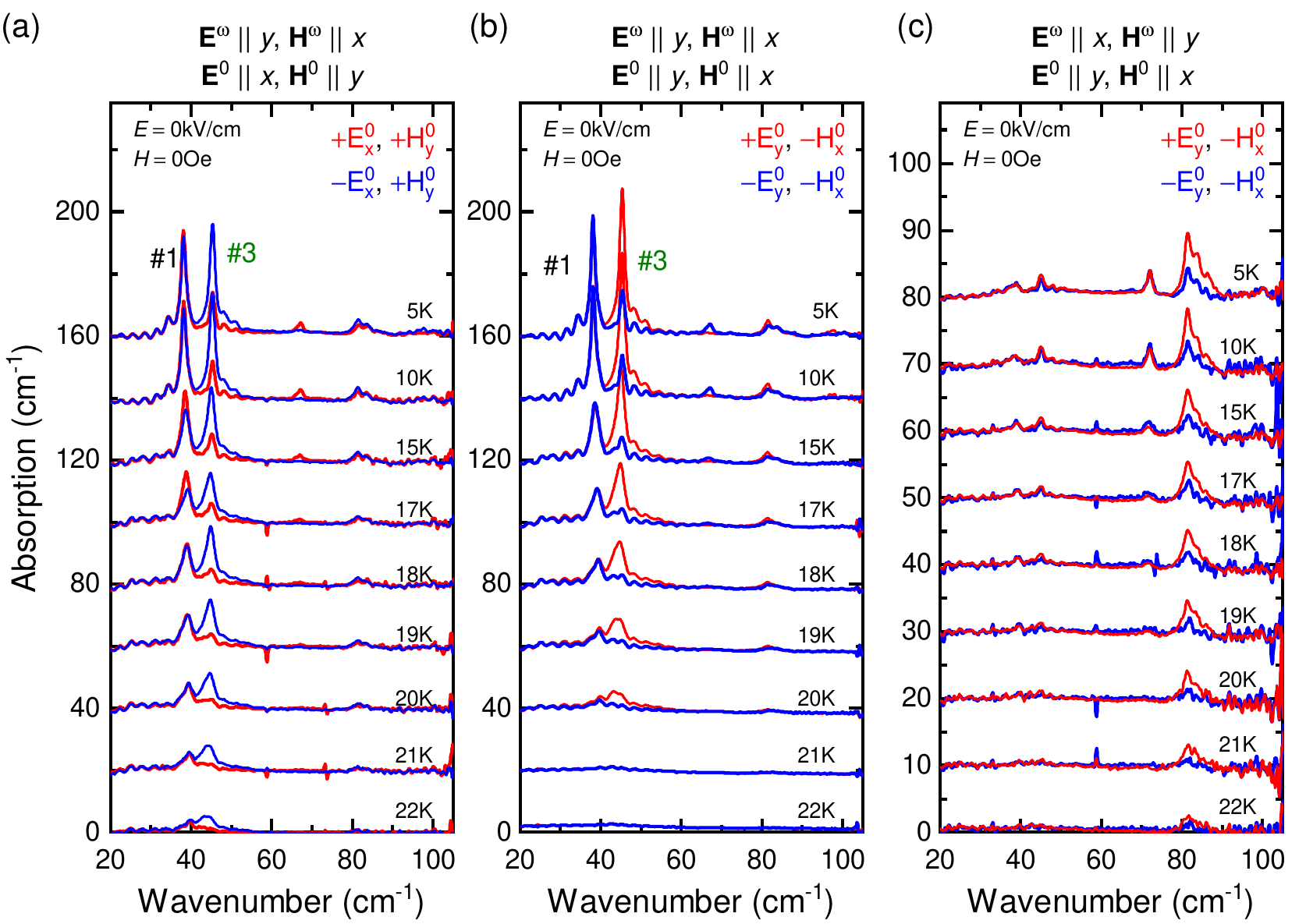}
    \caption{(Color online) Temperature dependence of the optical absorption spectra for the $(+E^0,+H^0)$ and $(-E^0,+H^0)$ poling fields, but in the absence of any external fields ($E$=0\,kV/cm, $H$=0\,Oe). Panels (a) to (c) correspond to the optical $\chi$ presented in Fig.~6(b). In each panel the optical spectra are shifted in proportion to the applied $H$ field.}
    \label{lcpoS02}
    \end{figure*}
    \end{center}
    %***********************************************************************************************************************************************

Magnetic field dependence of the magnetization measured at $T$=2\,K for fields applied along the main crystallographic axes are shown in Fig.~\ref{lcpoS01}(a).
In agreement with the earlier findings~\cite{Santoro1966,Rivera1994}, LiCoPO$_4$ has a antiferromagnetic (AFM) order with a magnetic easy-axis along the $y$ crystallographic direction.
For $\mathbf{H}\parallel{y}$, the AFM order is replaced at $H_{1,+}$=116\,kOe with an intermediate phase with $M^{\rm Sat}/3$ magnetization~\cite{Kharchenko2010}.
The partial spin-flop phase transition has hysteresis, the AFM phase is revived at $H_{1,-}$=112\,kOe, which refer to a first order phase transition.
Magnetic field induced polarization was measured at $T$=2\,K along the $x$ and $y$ axes for $H$ fields applied along the $y$ and $x$ axes, respectively (Fig.~\ref{lcpoS01}(b)).
The measurements were started from a single domain ME state, prepared by cooling the sample through the ordering temperature $T_{\rm N}$=21.3\,K in the presence of $E_x^0$=$+$2kV/cm and $H_y^0$=$+$100\,kOe or $E_y^0$=$+$2kV/cm and $H_x^0$=$+$140\,kOe poling fields.
Polarization was measured in the absence of $\mathbf{E}$ field during $\mathbf{H}$ field sweep (100\,Oe/s).
For $\mathbf{H}\parallel{x}$ the field induced polarization is linear up to $H$=140\,kOe with a slight decrease in the susceptibility.
For $\mathbf{H}\parallel{y}$, ferroelectric polarization disappears at the spin-flop phase transition and the ME order parameter does not fully recover when the field is lowered below $H_{1,-}$.
These findings are in excellent agreement with recent polarization measurements in pulsed magnetic field on crystals grown by flux method~\cite{Szewczyk2011,Khrustalyov2016}.

Figure~\ref{lcpoS02}(a) shows the optical absorption spectra measured at $T$=5\,K relative to the absorption spectra of the paramagnetic phase ($T$=30\,K) for six different configurations of the $E^\omega$ and $H^\omega$ of the light.
Each panels of the figure contain two spectra corresponding to the same light propagation direction.
Measurements were performed in the absence of external static fields, therefore each excitation has an absorption state corresponding to a multi-domain ME ground state.
Absorption spectra was taken between $k$=10\,cm$^{-1}$ and 150\,cm$^{-1}$. 
However, we found no spectral features between 10\,cm$^{-1}$ and 30\,cm$^{-1}$, while above $k$=120\,cm$^{-1}$ it was not possible to measure the optical absorption due to the strong absorption of the phonons.

\begin{center}
    \begin{figure*}[h]

    \includegraphics[width=17.0truecm]{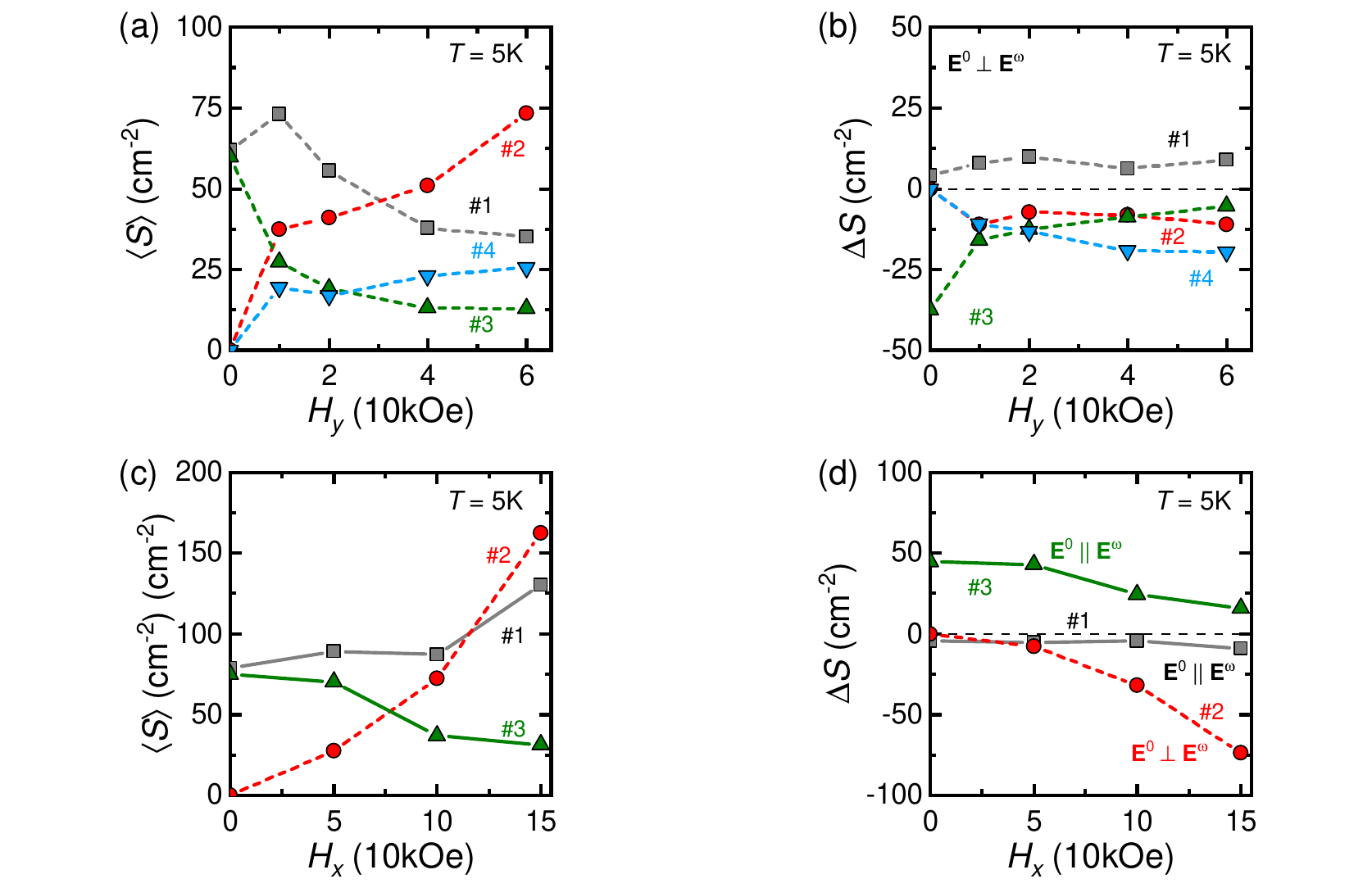}
    \caption{(Color online) Magnetic field dependence of the fitted $\left\langle{S}\right\rangle$ and $\Delta{S}$ values. Data in panels (a,b) correspond to the measurements in Fig.~4(a) with $\mathbf{E}^0\parallel{x}$ and $\mathbf{H}^0\parallel{y}$, while in panels (c,d) concludes Figs.~4(b) and 4(c) with $\mathbf{E}^0\parallel{y}$ and $\mathbf{H}^0\parallel{x}$. }
    \label{lcpoS03}
    \end{figure*}
    \end{center}

The absorption spectra were fitted with Lorentzian functions for further analysis:
\begin{equation}
\alpha(k)_{\pm} = \frac{2 S_{\pm}}{\pi} \frac{\gamma_{\pm}}{4(k-k_o)^2 + \gamma_{\pm}^2},
\end{equation}
where $\alpha(k)$ is the relative absorption spectra, $S_{\pm}$ are the oscillator strengths, $\gamma$ is the damping constant, and the $\pm$ signs correspond to the AFM-$\alpha$ and AFM-$\beta$ domains.
An averaged oscillator strength was calculated as the mean value of the oscillator strengths for the two polings according to $\left\langle{S}\right\rangle = (S_+ + S_-)/2$.
This quantity estimates the oscillator strength of the modes in the multi-domain case, \textit{i.e.} when the $\alpha$ and $\beta$ ME domains are equally populated, filtering out the effect of ODA.
The ODA, that is the absorption difference of between the $\alpha$ and $\beta$ ME domains is well estimated by the oscillator strength difference between the spectra measured with the to polings, $\Delta{S}= S_+ + S_-$.
For each mode observed, the field dependence of $\left\langle{S}\right\rangle$ is plotted in Fig.~\ref{lcpoS03} as a function of magnetic field.
The dynamic ME effect of the individual modes, presented in Fig.~5, correspond to $\Delta{S}$.
The sign difference between the $\Delta{S}$ and the $\chi_{xy}$ ME susceptibility of mode $\#$2 is attributed to Eq.~(1) of the main text.

    %***********************************************************************************************************************************************
    %SFIG #3
    %***********************************************************************************************************************************************

\end{document}